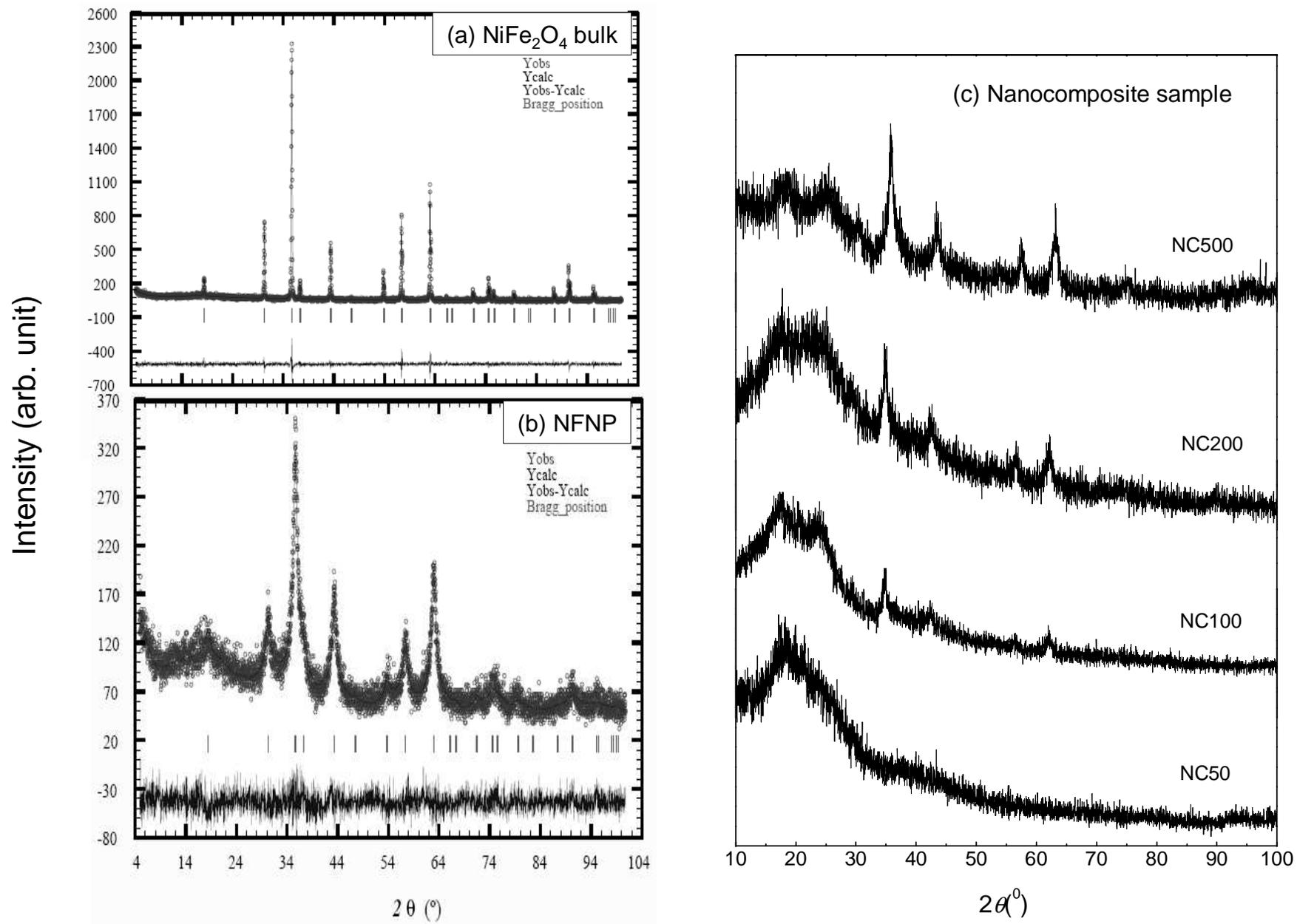

**Fig. 1**

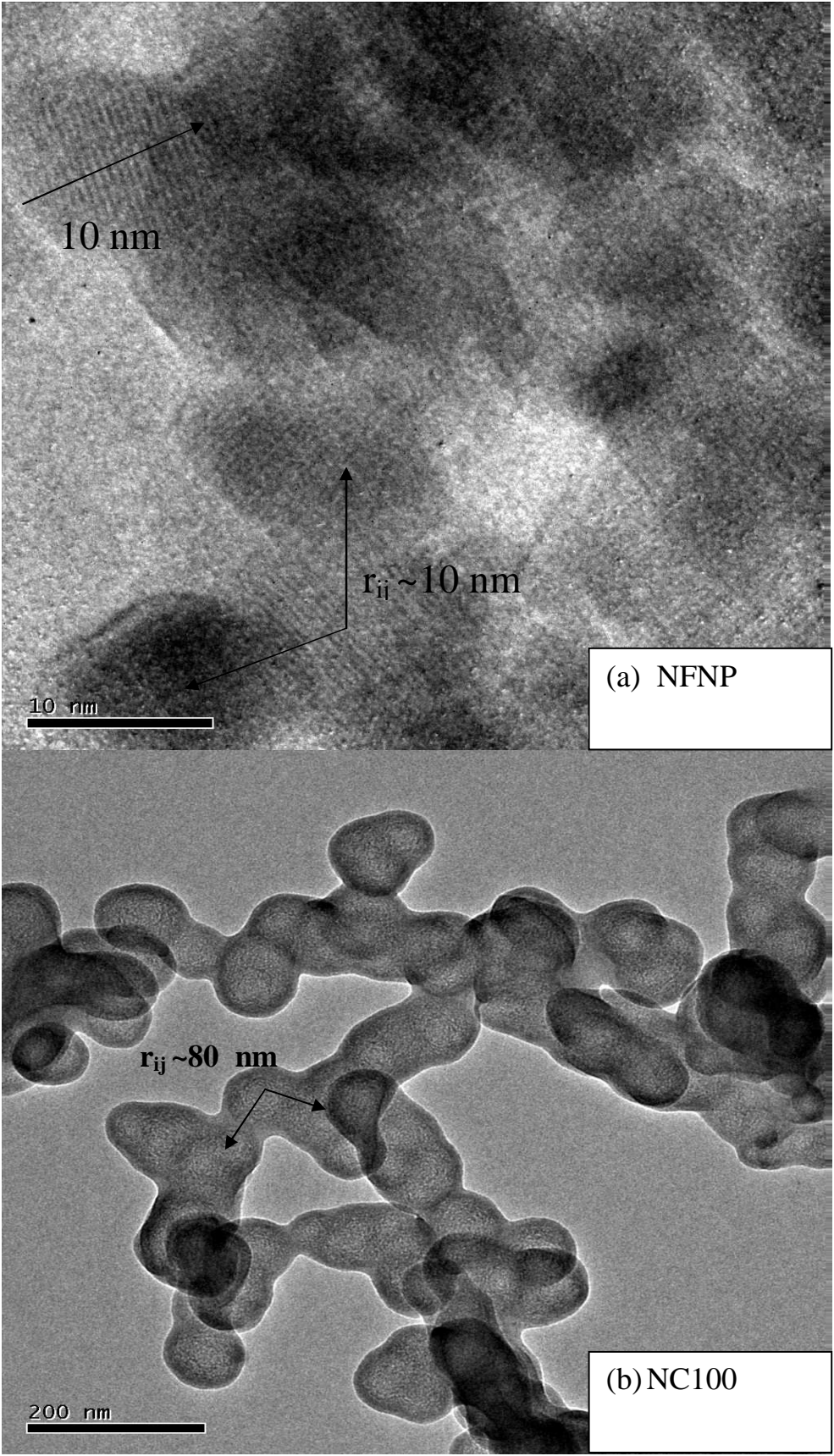

**Fig. 2**

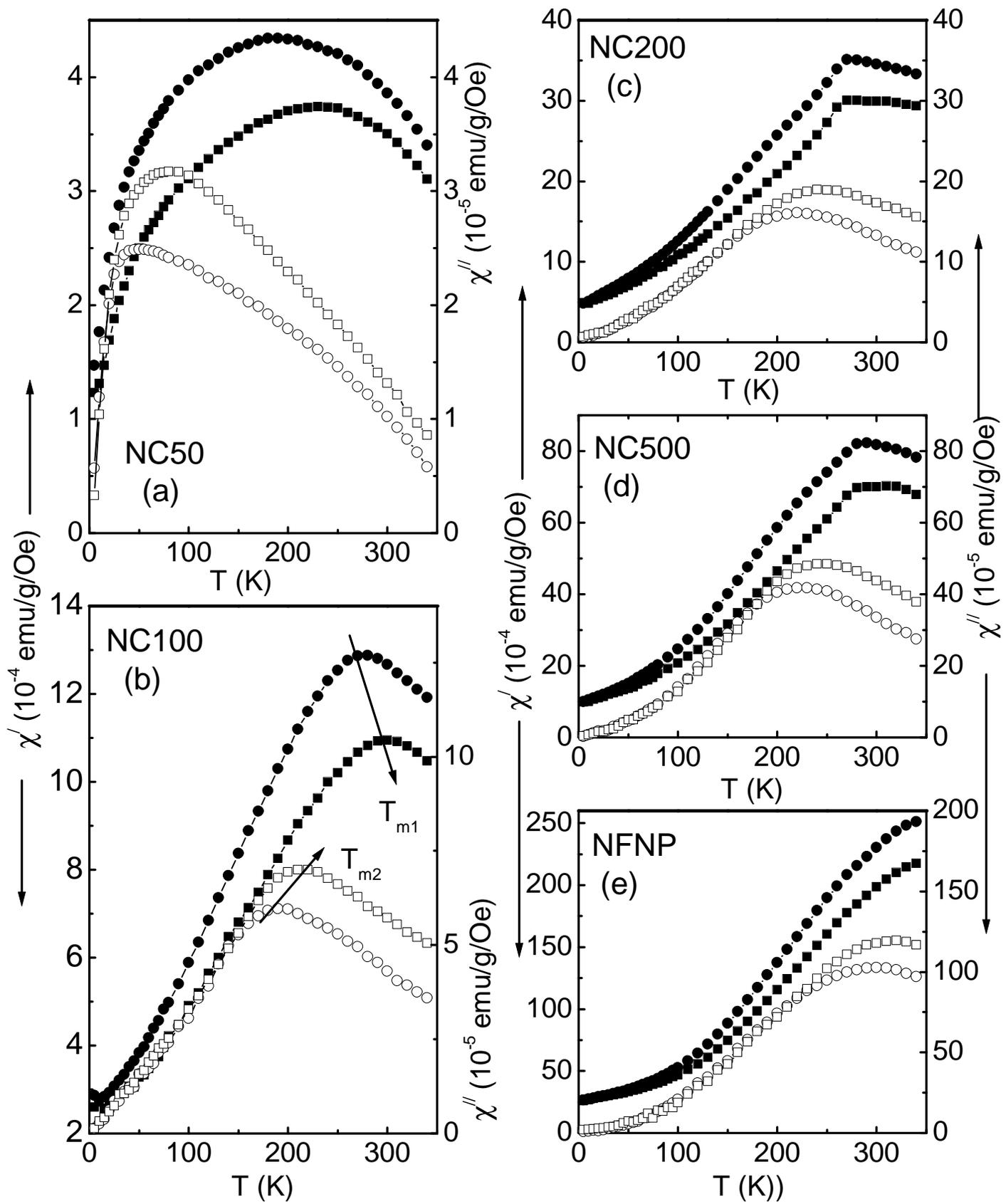

Fig. 3

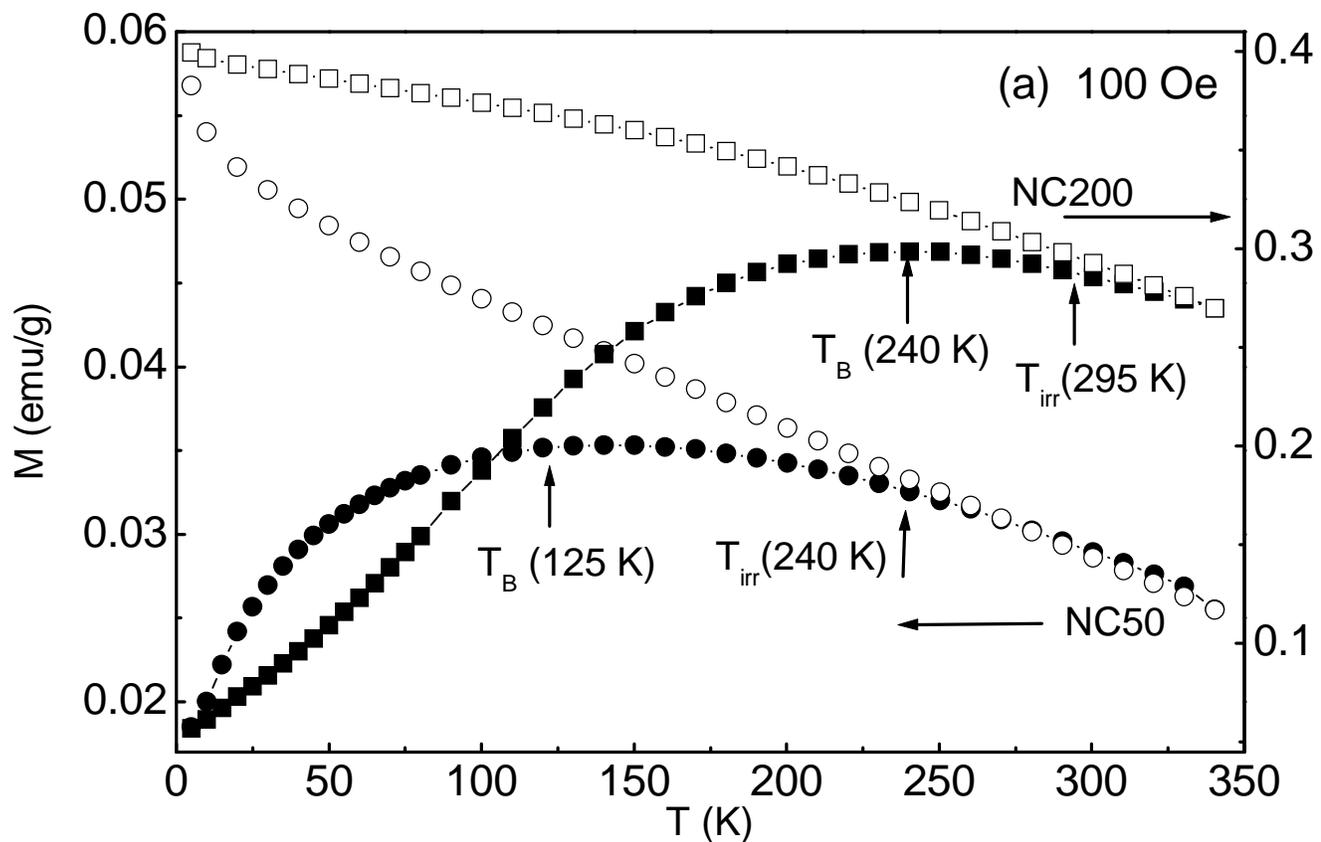
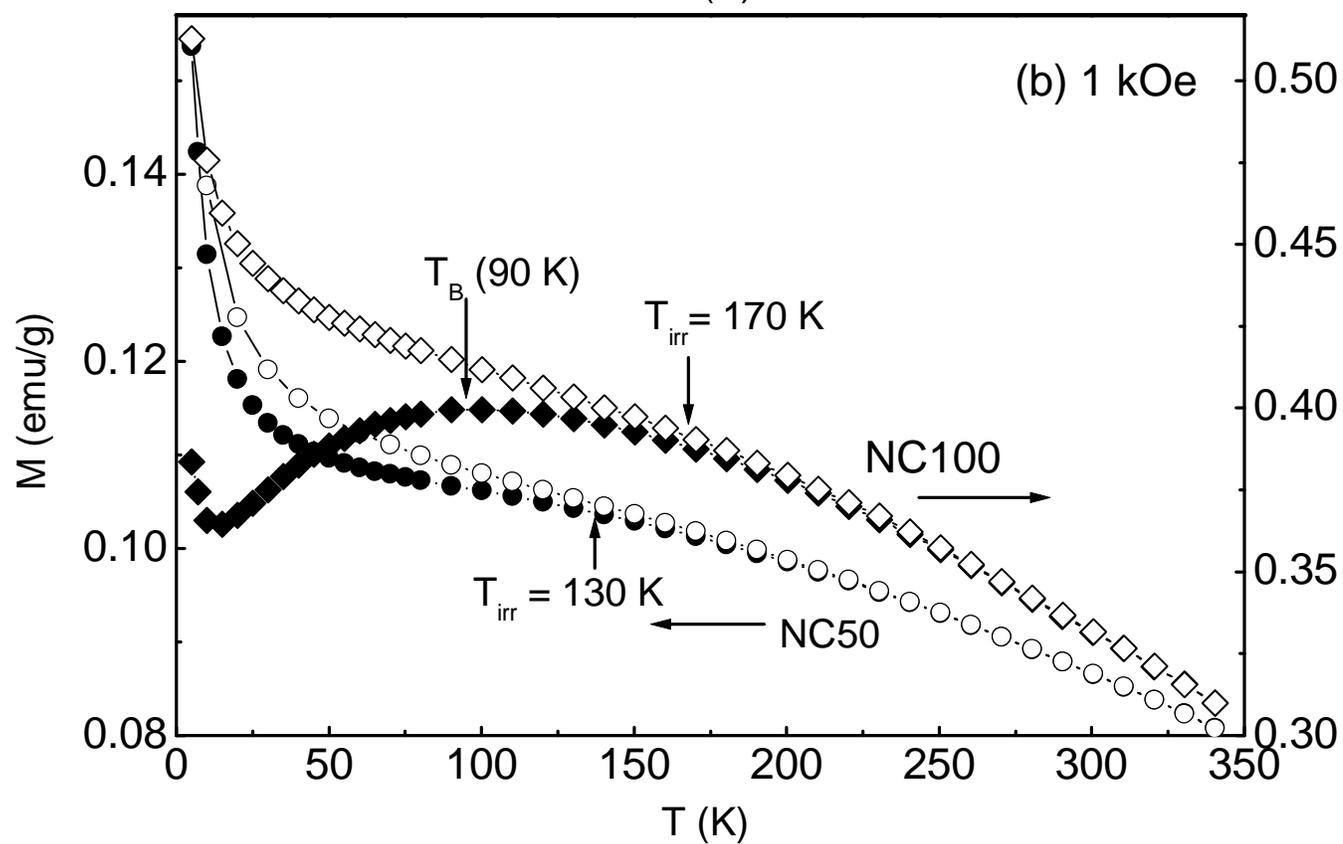

Fig. 4

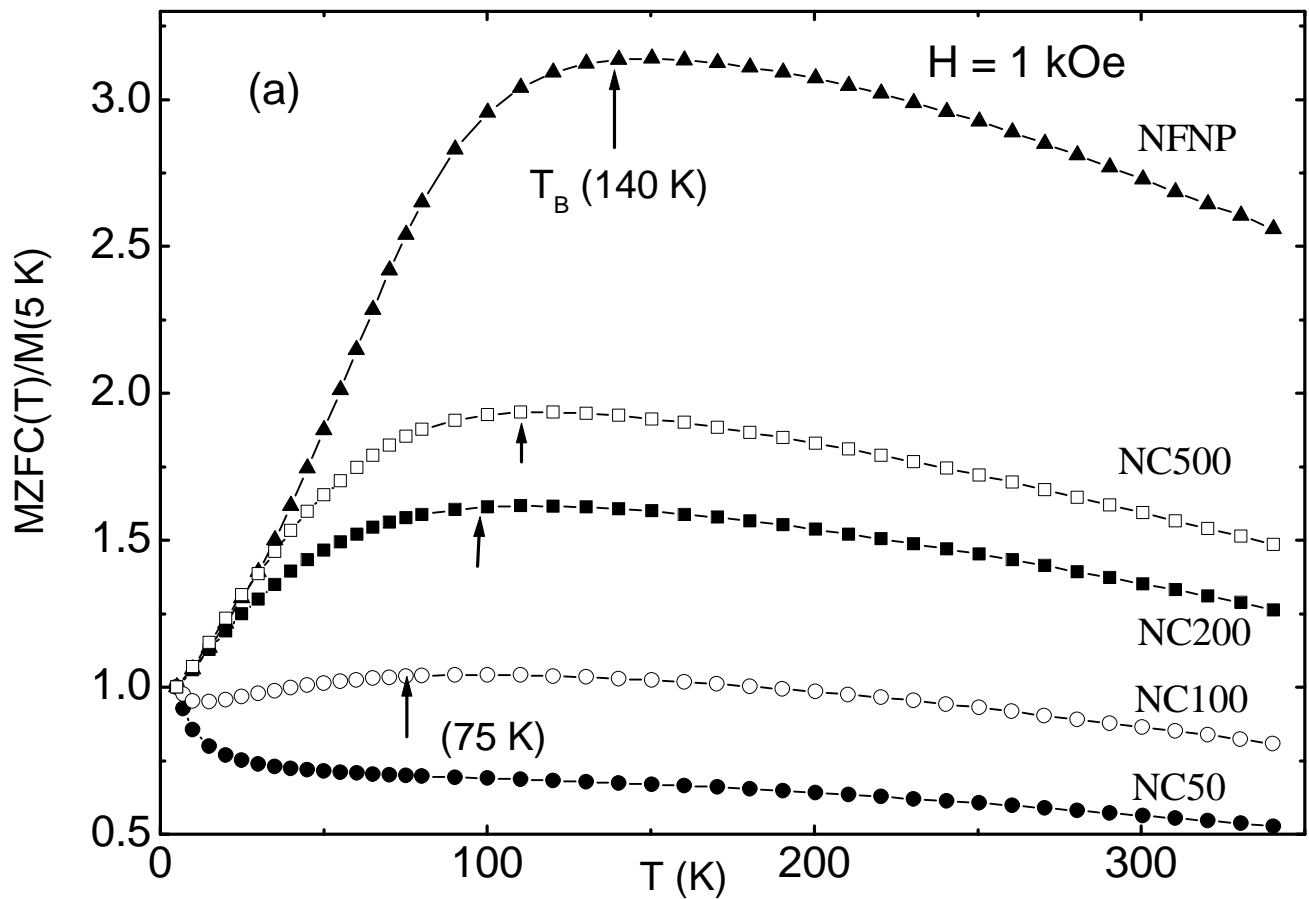
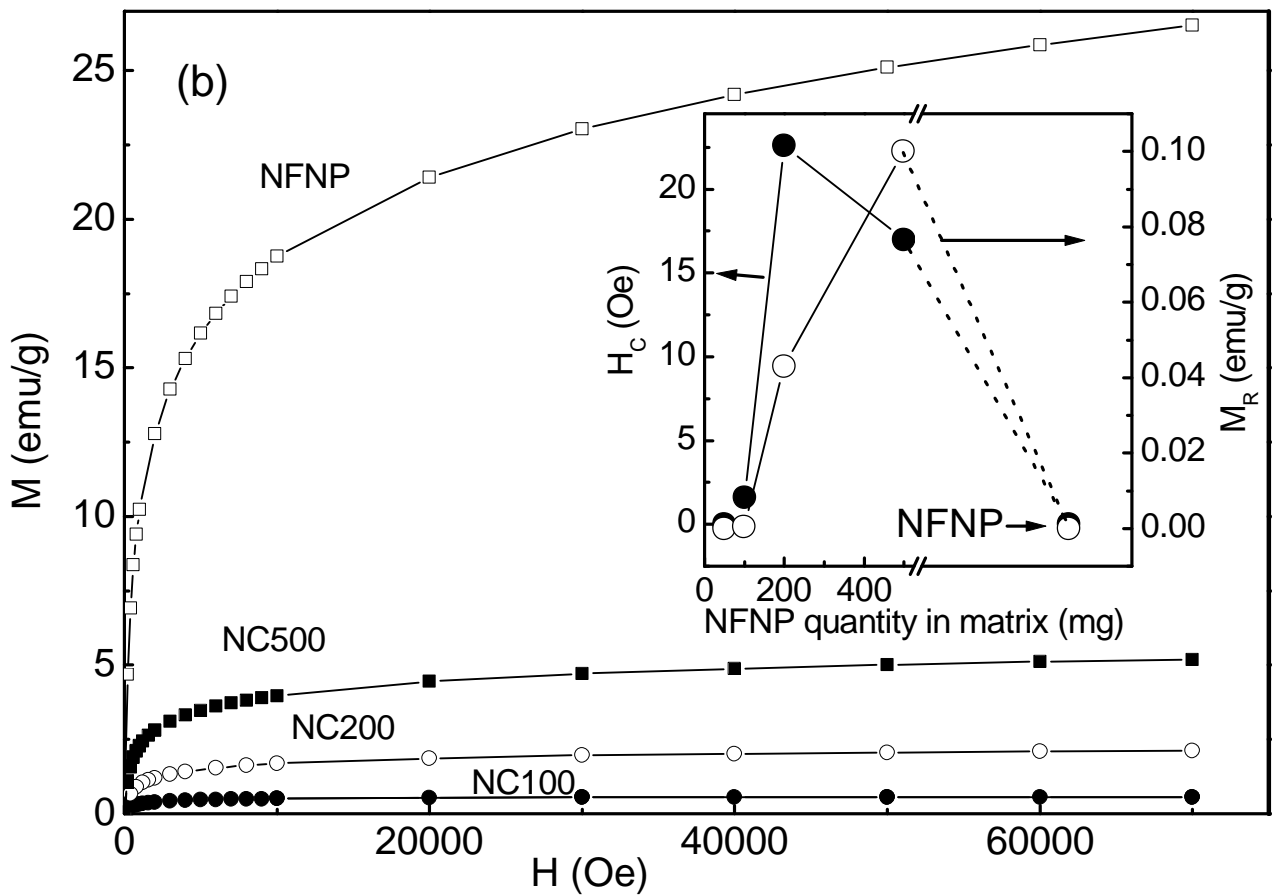

Fig. 5.

# Magnetic Response of $NiFe_2O_4$ nanoparticles in polymer matrix

A. Poddar[a], R.N. Bhowmik[b], Amitabha De[a] and Pintu Sen[c]

[a]*Saha Institute of Nuclear Physics, 1/AF Bidhannagar, Kolkata-700064, India*
[b]*Department of Physics, Pondicherry University, R. V.Nagar, Kalapet, Pondicherry-605014, India*
[c]*Variable Energy Cyclotron Centre, 1/AF Bidhannagar, Kolkata-700064, India*
E-mail address of the corresponding author(RNB): rnbhowmik.phy@pondiuni.edu.in

We report the magnetic properties of magnetic nano-composite, consisting of different quantity of $NiFe_2O_4$ nanoparticles in polymer matrix. The nanoparticles exhibited a typical magnetization blocking, which is sensitive on the variation of magnetic field, mode of zero field cooled/field cooled experiments and particle quantity in the matrix. The samples with lower particle quantity showed an upturn of magnetization down to 5 K, whereas the blocking of magnetization dominates at lower temperatures as the particle quantity increases in the polymer. We examine such magnetic behaviour in terms of the competitive magnetic ordering between core and surface spins of nanoparticles, taking into account the effect of inter-particle (dipole-dipole) interactions on nanoparticle magnetic dynamics.

# I. INTRODUCTION

The research activities on magnetic nanoparticles (MNPs) have seen many unusual phenomena [1, 2] over the last few decades. Some of the recently focused phenomena are: superparamagnetism, surface magnetism, spin glass, exchange bias effect, variation of particle magnetic moment and magnetic ordering temperature. The understanding of such magnetic phenomena has a direct or indirect importance on the theoretical modeling or applying the magnetic nanomaterials in technology. So many concurrent processes (inter-particle interactions, surface spin canting, interfacial/grain boundary effects, site exchange of cations among A and B sublattices, etc.) are involved in the magnetic properties of ferrite nanoparticles [3, 4, 5]. Most of the magnetic nanomaterials have shown superparamagnetic blocking of nanoparticles below a certain finite temperature. This behaviour of magnetic nanoparticles is believed to be a time domain problem, related to the inter-particle interactions and surface spin morphology of the particles. The superparamagnetic properties have been examined using techniques, like: magnetometry (dc magnetization, ac susceptibility) and Mössbauer spectroscopy [5, 6]. Kodama et al. [4] had proposed a core-shell model to explain the (ferri)magnetic properties of $NiFe_2O_4$ nanoparticles. It has been realized that the concept of core-shell spin structure may be the better description for nanoparticle magnetism [5, 7, 8]. Despite extensive work, the effect of core-shell spin structure and inter-particle interactions on the properties of magnetic nanoparticle is not clear. Recently, attempts are being made to study nano-composite materials, where magnetic nanoparticles are dispersed in a suitable (magnetic/non-magnetic) matrix [9, 10, 11]. The magnetic nano-composites are interesting not only to formulate the mechanism of inter-particle interactions and surface magnetism, but the materials are also emerging as the potential candidate to replace many conventional materials in science and technology [12, 13, 14].

In this paper, we study the magnetic nano-composite, involving $NiFe_2O_4$ nanoparticles in the polymer (3, 4-ethylenedioxythiophene) (**PEDOT**) matrix. The interesting fact is that $NiFe_2O_4$ is a typical ferrimagnet with relatively large magnetic moment (~55 emu/g) and ordering temperature (~850K) [15]. These magnetic parameters are important to synthesize nano-composite materials with reasonably large magnetic moment for room temperature applications. We have dispersed different quantity of $NiFe_2O_4$ nanoparticles (from same batch) in a fixed volume of PEDOT matrix. In this process, the complication related to the site exchange of cations, surface to volume ratio and size distribution as a function of varying particle size can be avoided.

## II. EXPERIMENTAL

$NiFe_2O_4$ nanoparticles are prepared by sol-gel procedure. The stoichiometric amount of $Fe(NO_3)_3$, $9H_2O$ and $Ni(NO_3)_2, 6H_2O$ are mixed (mass ratio 2:1) and dissolved in ethylene glycol at~ $40^0C$. The sol of metal salts is heated at ~ $60^0C$ to obtain gel. The gel product is dried at ~$100^0C$ and fired at ~ $400^0C$ for 24 hours. Finally, the sample is cooled to 300 K. The PEDOT nano-composite is prepared by polymerization of 3, 4-ethylenedioxythiophene (EDOT) in the colloidal dispersion containing specified quantity of $NiFe_2O_4$ nanoparticles (NFNP). Polymerization is allowed for 20 hours under vigorous stirring, resulting in a dark blue coloured nano-composite in the dispersed phase. Ethanol is added as a non-solvent to obtain the precipitate, which is washed and dried in vacuum. The nano-composites are denoted as NCX, where X indicates the quantity (50 mg, 100 mg, 200 mg, 500 mg) of $NiFe_2O_4$ nanoparticles dispersed in 3 ml volume of polymer. The pellet form of sample is used for characterization and measurement. The crystallographic phase of the samples is characterized by X-ray Diffraction spectrum at room temperature (300 K) using Cu $K_\alpha$ radiation from Philips PW1710 diffractometer. Particle size of the NFNP sample is determined from transmission electron micrographs (TEM). The ac susceptibility and dc magnetization of the samples are measured in the temperature range 5 K to 340 K, using SQUID magnetometer (Quantum Design, USA). The dc magnetization is recorded under zero field cooled (ZFC) and field cooled (FC) modes.

**III. RESULTS**

Fig. 1 (a-b) shows that the XRD spectrum of nanoparticle $NiFe_2O_4$ (NFNP) sample is consistent with the standard pattern of bulk $NiFe_2O_4$ sample. XRD patterns of both the samples are matched to cubic spinel structure with space group Fd3m. The crystal structure of the samples is determined by standard full profile fitting method using FULLPROF Program. The lattice parameter (*a*) is ~ 8.343 (2) Å and 8.351(3) Å for bulk sample and NFNP sample, respectively. The small variation of lattice parameter in NFNP sample with respect to the bulk sample may be attributed to the nanocrystalline nature of the material, characterized with broad peak lines. The particle size of NFNP sample (~ 7 nm) is estimated using Debye-Scherrer formula to (311) and (440) XRD peaks of the spectrum. The XRD pattern of NFNP sample is modified in the presence of polymer matrix (Fig. 1c). The nano-composite samples with lower particle quantity exhibit a broad background in XRD spectrum. The background is contributed from the amorphous polymer matrix [16]. The increase of particle quantity increases the crystalline fraction in the nano-composite, as seen from the appearance of crystalline peaks in NC500 sample. The TEM picture (Fig. 2a) suggests that NFNP particles are in nanocrystalline state with clear lattice fringe and size is ~ 10 nm. This value is close to the particle size (~ 7 nm) from XRD peaks. The TEM picture NC100 sample (Fig. 2b) shows that $NiFe_2O_4$ nanoparticles are coated by the amorphous matrix of polymer. The TEM picture suggested that particles are in contact for NFNP sample, while the particles are well separated in the nano-composite sample. The estimated inter-particle distance ($r_{ij}$) (~ 10 nm and 80 nm for NFNP and NC100 samples, respectively) is increasing with the decrease of particle quantity.

We, now, investigate the magnetic dynamics of nano-composites by measuring the ac susceptibility and dc magnetization. The real ($\chi'$) and imaginary ($\chi''$) parts of ac susceptibility data, measured at ac field ($h_{rms}$) ~ 1 Oe and frequency ($\nu$) = 10 Hz and 997 Hz, are shown in Fig. 3. The notable feature is that magnitude of susceptibility drastically decreases as the particle quantity decreases in the matrix. The samples showed a typical magnetic freezing/blocking behaviour below certain temperature. The freezing of $\chi'$ is appearing at higher temperature ($T_{m1}$) than the freezing of $\chi''$ at $T_{m2}$ ($T_{m2} < T_{m1}$), as shown by arrow in Fig. 3b for NC100 sample. The typical values of peak temperature ($T_{m1}$) may be noted as 180 K, 274 K at 10 Hz and 220 K, 300 K at 997 Hz for NC50 and NC100 samples, respectively. However, frequency dependent shift of $\chi'$ peak is not seen for NC200 and NC500 samples, except the onset of gradual decrease in $\chi'$ below 270 K and 280 K for NC200 and NC500 samples, respectively. The $\chi'$ data do not show any peak up to the measurement temperature 340 K for NFNP sample. It seems that $T_{m1}$ may be ~350 K for NFNP sample. On the other hand, the freezing temperature of $\chi''$ at $T_{m2}$ is clearly observed below 340 K for all samples and $T_{m2}$ strongly depends on the frequency of applied ac magnetic field. For example, $T_{m2}$ is ~ 40 K, 190 K, 218 K, 225 K, 300 K at 10 Hz and ~ 80 K, 216 K, 244 K, 248 K, 320 K for the samples NC50, NC100, NC200, NC500 and NFNP, respectively. The result clearly suggests that blocking/freezing temperature of the samples increases with the increase of particle quantity in the matrix. The estimated frequency shift per decade of frequency [$\Delta = \Delta T_{m2}/\{T_{m2}(10 \text{ Hz})\ln\nu\}$] is 0.220, 0.030, 0.024, 0.022 and 0.014 for the samples NC50, NC100, NC200, NC500 and NFNP, respectively. The decrease of $\Delta$ with increasing particle quantity in the present nano-composite system can be attributed to the increase of inter-particle interactions. However, the freezing behaviour can not be classified either an ideal spin glass or superparamagnetic blocking in the samples, because the typical value of $\Delta$ is ~ 1 for ideal superparamagnetic system and ~ 0.001 for spin glass system [17].

The dc field effect on the magnetic dynamics is understood from the measurement of zero field magnetization (MZFC) and field cooled magnetization (MFC) of the samples. The MZFC and MFC at 100 Oe for NC50 and NC200 samples are shown in Fig. 4a. The features suggest the blocking of magnetic nanoparticles, associated with the decrease of MZFC below the blocking temperature $T_B$ and separation between MFC and MZFC below the irreversibility temperature $T_{irr}$ ($\geq T_B$). Below $T_{irr}$, MFC of both the samples increases down to 5 K, except the increase is more rapid in NC50 than NC200 sample. It may be noted that $T_B$ ~125 K and 240 K at 100 Oe for NC50 and NC200 samples is less in comparison with $T_{m1}$ ~180 K and 270 K, as mentioned earlier from the $\chi'$ data. This is due to the effect of increasing dc field on the blocking temperature of particles [18]. The field effect is much more pronounced by increasing the field at 1 kOe (Fig. 4b). The novel feature is that MZFC of NC100 sample at 1 kOe shows blocking behaviour below $T_B$ ~80 K and an upturn in magnetization at lower temperature. The competition between low temperature upturn and superparamagnetic blocking about $T_B$ results in a minimum at ~ 20 K for the NC100 sample. Interestingly, a typical blocking behaviour in MZFC, as observed at 100 Oe (Fig. 4a), is not visible at 1 kOe for NC50 sample down to 5 K, except the observation of a low temperature upturn. The MFC (T) in both (NC50 and NC100) samples increases with similar manner below $T_{irr}$. However, the irreversible effect between MFC(T) and MZFC(T) curves at 1 kOe is significantly different in terms of splitting (between MFC and MZFC) and irreversibility temperature $T_{irr}$ (e.g.,~ 150 K and 200 K for NC50 and NC100 respectively). A systematic evolution of magnetization at 1 kOe with increasing particle quantity is seen from the MZFC(T) curves, normalized by 5 K data (Fig. 5a). The magnitude of MZFC at 5 K (~0.154, 0.384, 0.734, 1.431 and 3.727 in emu/g unit for NC50, NC100, NC200, NC500 and NFNP samples, respectively) increases with particle quantity. The interesting feature is that blocking temperature ($T_B$) decreases not only by the increment of field, but also by the decrease of particle quantity in the matrix. This is clear from the fact that $T_B$ ~140 K (at 1 kOe) of NFNP sample decreases to ~110 K (at 1 kOe) for NC500 sample, ~100 K (at 1 kOe) for NC200 sample, ~75 K (at 1 kOe) for NC100 sample and no typical blocking of magnetization at 1 kOe down to 5 K for NC50 sample. It is also noted that not MZFC alone, the MFC of NC50 and NC100 samples are also showing rapid increase at lower temperature. The low temperature upturn in MZFC is not seen for samples (NC200 and NC500) with sufficiently large quantity of magnetic nanoparticles. The magnetic moment of the nano-composite samples can be compared directly from the field dependence of magnetization (isotherms) data (Fig. 5b). The isotherms suggest that the typical ferrimagnetic character of $NiFe_2O_4$ nanoparticles are also retained in the nano-composite samples. The spontaneous magnetization can be estimated by the extrapolation of high field (> 40 kOe) magnetization data to zero field value. The obtained value of spontaneous magnetization is gradually increasing (~ 0.54 emu/g at 300 K for NC100, ~ 1.94 emu/g at 300 K for NC200, ~ 4.49 emu/g at 300 K for NC500 and ~ 21 emu/g at 300 K for NFNP) with the increase of particle quantity in the polymer matrix. This observation is also consistent with the variation of peak susceptibility at $T_{m1}$ (ac susceptibility data) and peak magnetization at $T_B$ (dc magnetization data) with the increase of particle quantity. It may be mentioned that NC50 sample exhibited a typical non-linear increase of M(H) data at 300 K, without any spontaneous magnetization. The M(H) variation of NC50 sample is not shown in Fig. 5 (b), because the magnitude is very low in comparison with other nano-composite samples. The interesting feature is that both coercivity ($H_C$) and remanent magnetization ($M_R$), obtained from M(H) data, are showing (inset of Fig. 5b) a maximum at the intermediate particle quantity, followed by zero value for both NC50 sample and NFNP sample (indicated by dotted lines). Similar variation of $H_C$ and $M_R$ with particle quantity in the matrix has been found in other nano-composite [12], where such magnetic properties have been attributed to the effect of dipolar inter-particle interactions.

## IV. DISCUSSION

We, now, try to understand some unique magnetic features of the present nano-composite samples. These features are reflected by a systematic variation in the shape of low temperature magnetization curves, where a typical blocking behaviour of magnetic particles is transformed into a magnetization upturn with the decrease of particle quantity in the matrix. The average blocking temperature ($T_B$) of the particles is defined at the peak/maximum of MZFC curves. The shift of magnetization maximum with increasing magnetic field is understood as an effect of magnetic field on the blocking of particle magnetization. In conventional superparamagnetic blocking, MZFC decreases with temperature below $T_B$, and one could expect either continuous increase or flatness of MFC below $T_B$ [19]. In our case, the shift of MZFC maximum is not due to increase of magnetic field alone, also due to the change of particle quantity in the matrix. The fact is that low temperature magnetic upturn is seen in both MZFC and MFC in our samples with lower particle quantity. The low temperature magnetization upturn has been found in a few composite systems [3, 20, 21]. Although various aspects, e.g., random anisotropy effect [18], surface spin contributions [3], spin reorientation [20] and precipitation of superparamagnetic type small particles [21], have been suggested, but the mechanism is not very clear till date. It is highly unexpected that the low temperature magnetic upturn is due to some precipitated small (superparamagnetic) particles in the samples, because the same NiFe$_2$O$_4$ nanoparticles with different quantity are mixed in the matrix. One could expect more precipitation in the samples with higher particle quantity. Consequently, low temperature upturn is expected to be more prominent for samples with higher particle quantity. However, this picture is not consistent with our experimental observations. We offer alternate explanation from the fact that similar low temperature magnetic behavior has been found in some antiferromagnetic nanoparticles [22, 23]. We demonstrate the origin of such low temperature magnetism in NiFe$_2$O$_4$ nano-composites in a simple and realistic manner, by incorporating the core-shell model proposed for antiferromagnetic nanoparticles (AFMNPs) [24] and considering the magnetic dipole-dipole interactions $E_{ij} = (\mu_0/4\pi)[\boldsymbol{\mu}_i.\boldsymbol{\mu}_j/r_{ij}^3 - 3(\boldsymbol{\mu}_i.\mathbf{r}_{ij})(\boldsymbol{\mu}_j.\mathbf{r}_{ij})/r_{ij}^5]$ [21]. The dipole-dipole interactions vary as a function of the magnitude of spin moments ($\boldsymbol{\mu}_i$ and $\boldsymbol{\mu}_j$), direction ($\theta_{ij}$) of the moments with respect to the line of joining ($\mathbf{r}_{ij}$) and the distance of separation ($r_{ij}$). The core spin moments in a typical AFMNP are compensated. The interactions between two spin moments (dipoles) may be from intra-particle or inter-particles. The inter-particles interactions, and also the core-shell interactions, are neglected in antiferromagnet due to small magnetic moment of particles. A significant number of uncompensated shell spins in the absence of strong exchange interactions contributes to the paramagnetic or superparamagnetic like magnetic upturn at lower temperatures, following the law [24]: M$_{ZFC}$ ∝ T$^{-\gamma}$ ($\gamma \leq 1$). The core-shell model for antiferromagnetic nanoparticle [24] has also found its application in ferrimagnetic nanoparticles [25]. The exchange interactions between core-shell spins are not neglected in ferrimagnetic (as in: NiFe$_2$O$_4$) nanoparticles due to large ferrimagnetic moment of core spins. The magnetic ordering of core spins dominates over the ordering of shell (surface) spins. Such magnetic competition strongly depends on the surface spin configuration and contributions from inter-particle interactions [9, 10, 11, 26]. J. Nogues et al [8] had described the essential role of shells in stabilizing the magnetism of core-shell nanoparticles, where the shell mediated interactions may introduce an induced magnetic state in the magnetic nanoparticle. Our experimental results suggested that the ferrimagnetic character of $NiFe_2O_4$ nanoparticles is maintained in the nano-composites. We assume that the change of surface morphology for each particle, due to pinning of spins to the polymer bonds [27], remains identical as long as $NiFe_2O_4$ nanoparticles are concerned in the PEDOT matrix. We, further, assume that the surface spin ordering (surface spin morphology) depends on the effective field, created by the exchange interactions between shell-core spins (intra-particle interactions) and the interactions between neighbouring particles (inter-particle interactions). It may be mentioned that change of exchange interactions of the particles are not playing significant role with the change of particle quantity in the matrix, because distortion in exchange interactions occur only between core and shell spins. The interactions between two particles are included in the form of dipole-dipole interactions and this effect is significant with the increasing particle quantity in the matrix. The increase of inter-particle interactions with particle quantity is realized from a systematic decrease of the shift of ac susceptibility maximum per decade of frequency ($\Delta$).

Now, we estimate the relative strengths of dipole-dipole interactions from the variation of inter-particle separation ($r_{ij}$) ~ 10 nm and ~ 80 nm (from TEM picture) and magnetic moment ($\mu$) at 300 K ~ 0.54 emu/g and ~ 21 emu/g for NFNP and NC100 samples (from M-H data), respectively. We simplify the dipole-dipole interaction term as $E_{ij} \sim -(2\mu_0/4\pi)\mu_i\mu_j \cos\theta_{ij}/r_{ij}^3 \sim -(2\mu_0/4\pi)\mu^2 \cos\theta_{ij}/r_{ij}^3$. The ratio of dipole-dipole interactions for NFNP and NC100 samples is $(E_{ij})_{NFNP}/(E_{ij})_{NC100} \sim (\mu_{NFNP}/\mu_{NC100})^2 (r_{NC100}/r_{NFNP})^3 (\cos\theta_{NFNP}/\cos\theta_{NC100})^2$. Replacing $(E_{ij})_{NFNP}/(E_{ij})_{NC100}$ by $(T_{m1})_{NFNP}/(T_{m1})_{NC100}$ (from $\chi'$ peak temperature at 10 Hz) ~ 350 K/274 K and substituting all the parameters values, we get $\cos\theta_{NFNP}/\cos\theta_{NC100}$ ~0.00128. This suggests that the angle ($\theta_{ij}$) between two dipole moments of NC100 sample is less than that for the NFNP sample. The $\theta_{ij}$ is restricted to

$0^0 < \theta_{ij} < 90^0$. The lower value of $\theta_{ij}$ for the samples with lower particle quantity is explained from the fact that dipole-dipole interactions are not large enough, due to large inter-particle separation ($r_{ij}$), to significantly modify the surface spin ordering. The individual particle moments, largely controlled by the superparamagnetic type contributions from shell spins, are relatively free to respond to the external magnetic field. The superparamagnetic nature (less inter-particle interactions) in NC50 samples is suggested from a typical M(H) curve at 300 K with zero coercivity. As soon as the small inter-particle interactions are minimized, the superparamagnetic response of the shell spins dominates in controlling the low temperature magnetic behaviour of the material. The increase of particle concentration in the matrix contributes to a significant amount of dipole-dipole interactions [3, 9, 11, 28], which effectively increase the anisotropy energy ($E_{eff} \propto T_B$) of the nanoparticles. The particle moments are randomly blocked along the local anisotropy axes, created by the inter-particle (dipole-dipole) interactions from the neighbouring particles. So, the magnetic behaviour is no more a surface dominant effect, rather the particles moments are collectively blocked below the average blocking temperature ($T_B$) of the sample.

## V. CONCLUSIONS

The experimental data indicated that the shape of low temperature magnetization curves and magnetization blocking of $NiFe_2O_4$ nanoparticles in PEDOT polymer matrix depends not only on the factors like: particle size and core-shell morphology alone, but also on the factors like: variation of magnetic field, variation of particle quantity in the matrix and mode of magnetization measurement. The correlated effect of core-shell spin structure and inter-particle interactions are used to understand the shape of magnetization curves. The paramagnetic contributions of shell spins are significant for the samples with lower particle quantity in the matrix. The inter-particle interactions, contributed by dipole-dipole interactions between particle moments, are dominating in samples with higher particle quantity. The magnetization upturn at lower temperatures or the magnetization maximum at finite temperature is an effect of modified surface spins dynamics, depending on the quantity of magnetic particles in the polymer matrix.

**Acknowledgment:** The authors thank Pulak Ray for providing TEM data.

**References:**

[1] J L Dormann, L Bessais and D Fiorani, J. Phys. C: Solid State Phys. **21**, 2015 (1988)

[2] J. Nogués et al., Physics Reports **422**, 65 (2005)

[3] E. Tronc et al., J. Magn. Magn. Mater. **221**, 63 (2000)

[4] R.H. Kodama, A.E. Berkowitz, E.J. McNiff Jr., S. Foner , Phys. Rev. Lett. **77,** 394 (1996).

[5] V. Sepelak, I. Bergmann, A. Feldhoff, P. Heitjans, F. Krumeich, D. Menzel, F.J. Litterst, S.J. Campbell and K. D. Becker, Chem. Mater**. 111**, 5026  (2007).

[6] T. Jonsson, P. Nordblad, and P. Svedlindh, Phys. Rev. B **57,** 497 (1998)

[7] Òscar Iglesias, Xavier Batlle, and Amílcar Labarta, Phys. Rev. B **72**, 212401 (2005)

[8] J. Nogues, V. Skumryev, J. Sort, S. Stoyanov, and D. Givord, Phys. Rev. Lett. **97**, 1572003 (2006)

[9] J.L. Dormann et al., Phys. Rev. B **53**, 14291 (1996)

[10] J. Zhang, C. Boyd, and W. Luo, Phys. Rev. Lett. **77**, 390 (1996)

[11] C. Djurberg, P. Svedlindh and P. Nordblad, Phys. Rev. Lett. **79**, 5154 (1997)

[12]  A. Ceylan, C.C. Baker, S.K. Hasanain,, S.I. Shah,  Phys. Rev. B, **72**, 134411 (2005)

[13] R.F. Ziolo  et al., Science **257**, 219 (1992)

[14] J. L. Wilson, P. Poddar, N. A. Frey, H. Srikanth, K. Mohomed, J. P. Harmon, S. Kotha and J. Wachsmuth, J. Appl. Phys. **95**, 1439 (2004)

[15] U. Lüders, M. Bibes, Jean-François Bobo, M. Cantoni, R. Bertacco, and J. Fontcuberta, Phys. Rev. B **71**, 134419 (2005)

[16] P. Dallas, V. Georgakilas, D. Niarchos, P. Komninou, T. Kehagias and D. Petridis, Nanotechnology **17,** 2046 (2006)

[17] J.A. Mydosh, Spin Glasses: an Experimental Introduction, Taylor & Francis, London, 1993

[18] W. Luo, S.R. Nagel, T. F. Rosenbaum, and R. E. Rosensweig, Phys. Rev. Lett. **67**, 2721 (1991)

[19] Qi Chen and Z.J. Zhang, Appl. Phys. Lett. **73**, 3156 (1998)

[20] R.V. Upadhyay, K. Parekh, L. Belova, K.V. Rao, J. Magn.Magn. Mater. **311**, 106 (2007)

[21] R.D. Zysler, C.A. Ramos, E. De Biasi, H. Romero, A. Ortega, D. Fiorani, J.Magn. Magn. Mater. **221**, 37 (2000)

[22] L. He and C. Chen, N. Wang, W. Zhou, and L. Guo, J. Appl. Phys. **102**, 103911 (2007)

[23] A. Punnoose, H. Magnone, M.S. Seehra, J. Bonevich, Phys. Rev. B **64,** 174420 (2001).

[24] R.N. Bhowmik, R. Nagarajan, and R. Ranganathan, Phys. Rev. B **69**, 054430 (2004).

[25] T. Zhang, T. F. Zhou, T. Qian, and X. G. Li, Phys. Rev. B **76**, 174415 (2007)

[26] P. Poddar, J.L. Wilson, H Srikanth, S.A. Morrison and E.E. Carpenter, Nanotechnology **15,** S570 (2004)

[27] S.R. Ahmed, S.B. Ogale, G.C. Papaefthymiou, R. Ramesh, P. Kofinasa, Appl. Phys. Letter. **80**, 1616 (2002)

[28] M. Blanco-Manteco´n, K. O'Grady, J. Magn. Magn. Mater. **296**, 124 (2006)

**Figure Captions:**

Fig. 1 (a) XRD spectrum for bulk (a) and nanoparticle (b) NiFe2O4 samples, alongwith Full profile fit data. The XRD spectrum of nanocomposite samples are shown in (c).

Fig. 2. The TEM pictures are shown for NFNP sample (a) and NC100 sample (b). The arrow indicates the average inter-particle separation ($r_{ij}$) of the particles.

Fig. 3. Real $\chi'$ (left scale) and imaginary $\chi''$ (right scale) parts of ac susceptibility data for different nano-composite samples, measured at $h_{rms}$ = 1 Oe and $\nu$ = 10 Hz (circle) and 997 Hz (square). Solid and Open symbols represent $\chi'$ and $\chi''$, respectively. The arrows (Fig. 2b) indicate the shift of $\chi'$ and $\chi''$ maximum with increasing frequency.

Fig. 4. The temperature dependence of MZFC and MFC data at 100 Oe for NC50 and NC200 samples in (a) and at 1 kOe for NC50 and NC100 samples in (b). $T_B$ and $T_{irr}$ of the samples are indicated by (up-down) arrows. The side arrows represent the magnetization axis for the corresponding samples.

Fig. 5. (a) MZFC (normalized by 5 K data) at 1 kOe for different nano-composite samples. The arrows indicate the position of blocking temperature ($T_B$) of the samples. (b) Field dependence of magnetization at 300 K for the selected samples. Inset of Fig. 5(b) shows the variation of coercive field ($H_C$) and remanent magnetization ($M_R$) with particle quantity in the matrix. The dotted lines join the $H_C$ and $M_R$ of NC500 sample with NFNP sample.